\documentclass[12pt,a4paper]{article}
\usepackage{cite}
\usepackage{dsfont}
\usepackage{latexsym}
\usepackage{epsfig}
\usepackage{amsfonts}
\usepackage{array,longtable}
\usepackage{graphicx}
\usepackage{amssymb}

\newcommand{\beq}{\begin{equation}} \newcommand{\eeq}{\end{equation}}
\newcommand{\beqa}{\begin{eqnarray}} \newcommand{\eeqa}{\end{eqnarray}}
\newcommand{\mueg}{$\mu \to e \gamma$}

\begin{document}

\begin{center}

\vspace{2cm}

{\bf \Large {\LARGE $\mu\to e\gamma$} Decay Rate in the MSSM with\\ [0.5cm] Minimal Flavour Violation}

\vspace{2cm}

{\bf \large M. Davidkov$^{1}$ and D. I. Kazakov$^{1,2}$}\vspace{0.5cm}

{\it $^1$Bogoliubov Laboratory of Theoretical Physics, Joint
Institute for Nuclear Research, Dubna, Russia, \\
$^2$Institute for Theoretical and Experimental Physics, Moscow, Russia}
\vspace{1cm}

\abstract{The branching ratio for the   \mueg\  decay in the framework of the minimal flavour violation in the MSSM is calculated
for various regions of the MSSM parameter space.  The lepton flavour violation goes through the PMNS mixing matrix. The dependence on $\tan\beta$ is studied in comparison with experimental data. The results crucially depend on the  mixing angle
$\theta_{13}$.  Observation of this decay would serve as a manifestation of new physics beyond the SM.}

\end{center}

\section{Introduction}

The lepton flavour violating (LFV) processes are forbidden in the Standard Model  if all neutrinos are massless.
They  neither go via the  charged currents nor via neutral ones.  The situation changes if neutrinos are massive.
Then one has to introduce the right-handed  neutrinos to the SM and the lepton sector becomes similar to the quark one.
By analogy with the CKM mixing matrix one obtains the PMNS mixing matrix whose
 elements are measured today with
a reasonable accuracy.  The LFV processes  become possible via the charge currents. FCNC are still forbidden.

The  $\mu \to e\gamma$  decay  is one of the most important LFV processes. In the SM with neutrino masses it
is described by the so-called penguin diagram and is proportional to the off-diagonal elements of the PMNS matrix.
Due to chirality conservation it is also proportional to the neutrino masses. Calculations in the SM give \cite{SM-Br1, SM-Br2, SM-Br3, SM-Br4, SM-Br5}
\beqa
Br_\mathrm{SM}\left(\mu \to e \gamma\right)= \frac{3 \alpha}{32\pi}\left| \frac{\Delta m^2_{21}}{m_W^2}U_{e2}^* U_{\mu 2} +  \frac{\Delta m^2_{31}}{m_W^2}U_{e3}^* U_{\mu 3} \right|^2
\label{SM-Br}
\eeqa
With the present neutrino oscillation data one can estimate the branching ratio $Br_{\mathrm{SM}}( \mu \to e\gamma) \sim 10^{-55}$, a ridiculously small number.
So the only reasonable possibility to observe such a decay would be a non-minimal flavour violation or a new physics
with heavy particles propagating in the loops.
Modern experiments \cite{Adam:2009ci} have reached the upper bound of
$$Br(\mu\to e \gamma)\leq 2.8\cdot10^{-11},$$
and the MEGA experiment \cite{Brooks:1999pu} 
 gives a similar result
$$Br(\mu\to e\gamma)\leq1.2\cdot10^{-11}.$$ They observe some "suspicious" events, though of no statistical value so far.

One sees that the experimental bound is far from the SM prediction. However, in extensions of the SM, for instance, in the Minimal Supersymmetric Standard Model (MSSM) there are additional  contributions to the LFV processes. In particular, one can easily get
the branching ratio for the $\mu \to e\gamma$  decay in the MSSM of the order of $10^{-11}$, which makes the measurement of this
decay rate extremely sensitive to new physics. One can say that  if the decay is observed this would be the manifestation of physics beyond the SM.


\section{Mixing matrix in the lepton sector}

When neutrinos are massive, the lepton sector of the SM resembles the quark one and naturally includes the mixing of flavours. The left-handed fields of the neutrino flavour eigenstates $\nu_{e_L}$, $\nu_{\mu_L}$, $\nu_{\tau_L}$ are linear combinations of the three neutrino mass eigenstates $\nu_{1_L}$, $\nu_{2_L}$, $\nu_{3_L}$
$$\nu_{l_L} = \sum_{i=1}^3 U_{li}\nu_{i_L}$$
and the weak interaction via the charged lepton current  reads
\beq
\mathcal{L}_{CC} = -\frac{g_2}{\sqrt 2}\sum_{l=e, \mu, \tau}\overline{l}_L\gamma_\alpha \nu_{l_L} W^{\alpha \dagger} + h.\,c.,
\eeq
where 
\beqa
U =
\left(
\begin{array}{ccc}
U_{e1} & U_{e2} & U_{e3} \\
U_{\mu 1} & U_{\mu 2} & U_{\mu 3} \\
U_{\tau 1} & U_{\tau 2} & U_{\tau 3} 
\end{array}
\right)
\eeqa
is the unitary neutrino mixing matrix, the so-called PMNS matrix \cite{PMNS1, PMNS2}.
The PMNS matrix can be parametrised by three angles and, depending on whether the massive neutrinos are Dirac or Majorana particles, by one or three CP violating phases \cite{Bilenky, Schechter, Doi}. We write
\beqa
U =
\left(
\begin{array}{ccc}
{c_{12}} {c_{13}} & {s_{12}} {c_{13}} & {s_{13} e^{i\delta}} \\
-{s_{12}} {c_{23}}-{c_{12}} {s_{13}} {s_{23}}e^{i {\delta}} &  {c_{12}} {c_{23}}- {s_{12}} {s_{13}} {s_{23}}e^{i {\delta}} & {c_{13}} {s_{23}} \\
{s_{12}} {s_{23}}-{c_{12}}{s_{13}} {c_{23}} e^{i {\delta}} & -{c_{12}} {s_{23}} -{c_{23}} {s_{12}} {s_{13}} e^{i {\delta}} & {c_{13}} {c_{23}}
\end{array}\right)T
\eeqa
where $c_{ij}\equiv\cos\theta_{ij}$, $s_{ij}\equiv\sin\theta_{ij}$, $\delta$ is the Dirac CP violating phase, and the matrix
\beq
T=\mathrm{diag}\left(1, e^{i\frac{\alpha_{21}}{2}},e^{i\frac{\alpha_{31}}{2}}\right)
\eeq
contains the Majorana CP violating phases $\alpha_{21}$ and $\alpha_{31}$.
In the following, we neglect all CP violating phases assuming CP conservation in the lepton sector.  
The existing experimental neutrino oscillation data allow the determination of the solar and atmospheric neutrino oscillation parameters $\theta_{12}$ and $\theta_{23}$ with a relatively good precision. It is also possible to place rather stringent bounds on the angle $\theta_{13}$. By analysing the experimental data it has been found that \cite{Fogli:2008ig, Schwetz:2008er}
$$\sin^2 \theta_{12} = 0.304^{+0.022}_{-0.016}\,, ~~~\sin^2 \theta_{23}=0.5^{+0.07}_{-0.06}.$$  
Even though the angles $\theta_{12}$ and $\theta_{23}$ are known with a reasonable accuracy, the angle $\theta_{13}$ still remains unknown and is the main source of uncertainties of our predictions. A combined 3-neutrino oscillation analysis of the global data gives \cite{Schwetz:2008er}
$$ \sin^2 \theta_{13} \leq 0.035~(0.056)~~\mathrm{at}~90\%~(99.73\%)~\mathrm{C.L.} $$ and a global analysis of all available neutrino oscillation data provides the numerical value \cite{Fogli:2008jx} $$\sin^2 \theta_{13}= 0.016 \pm 0.010.$$ 
As it will be clear later, the $\mu\to e\gamma$ decay rate is proportional to $\sin\theta_{13}$ and vanishes with the latter.

\section{The $\mu\to e\gamma$ decay rate}  

\begin{figure}[htb]\vspace{0.3cm}
\begin{center}
 \epsfxsize=4.cm
 \epsffile{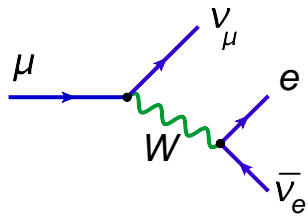}
 \caption{The double neutrino muon decay}
 \label{mu}
 \end{center}
 \end{figure}

The main decay of muon is the double neutrino decay $\mu\to e\nu_\mu\bar{\nu}_e$ which gives almost 100\% of the width and is one of the best measured decays. In the SM  it goes through the $W$-boson exchange (see Fig.\ref{mu}) and the decay width is given by
\beq\label{wid}
\Gamma(\mu \to e \bar\nu \nu) = \frac{G_F^2 m_\mu^5}{192 \pi^3}\,F\left( \frac{m_e^2}{m_\mu^2}\right),
\eeq
where
\beq
F(x) = 1-8x-12x^2 \ln x + 8x^3-x^4
\label{Fx}
\eeq
accumulates the radiative corrections and the Fermi constant is related to the weak coupling and the $W$-mass
by $G_F={g^2}/{4\sqrt{2}m_W^2}$.
Since  $m_e \ll m_{\mu}$, the value of $F(m_e^2/m_{\mu}^2) = 0.9998$ is very close to 1.

The vertices in Fig.\ref{mu} contain the PMNS matrix elements. In the case when one does not distinguish between neutrino flavours, one has to sum over all neutrino species  and due to absence of interference obtains the sums $\sum_i U_{\mu i}U^*_{\mu i}=1$ and  $\sum_i U_{e i}U^*_{e i}=1$. The latter are a consequence of the unitarity of the PMNS matrix. So the decay width (\ref{wid}) coincides with the one obtained in the SM without neutrino mixing. On the contrary, if one distinguishes neutrino flavours, the decay rate is proportional to specific PMNS matrix elements and is smaller.

\begin{figure}[htb]\vspace{0.3cm}
\begin{center}
 \epsfxsize=15.cm
 \epsffile{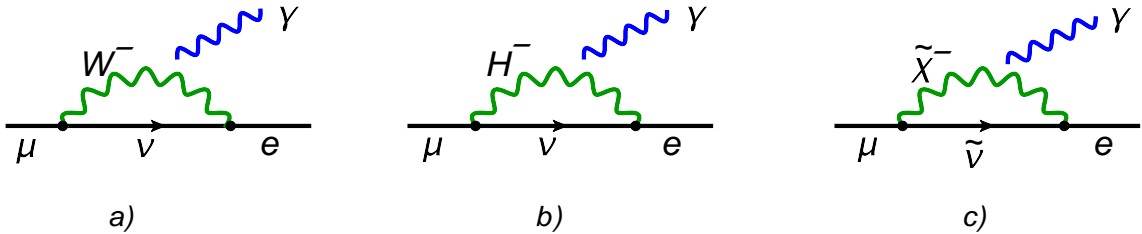}
 \caption{The penguin diagrams contributing to the $\mu\to e\gamma$ decay in the SM  (a) and in the MSSM (a, b, c).
 Tilde denotes the superpartner of the corresponding particle.}
 \label{1}
 \end{center}
 \end{figure}

Considering only the main contribution stemming from the electromagnetic penguin operator the \mueg\ transition amplitude is given by \cite{Hisano:1995cp, Hisano:1996qq, Okada:1999zk, Carvalho:2001ex, Hisano:1995nq}
\beq 
\Gamma(\mu\to e \gamma)=
\frac{G_F^2 m_\mu^5\alpha}{32\pi^4}\,
\left|A_{W^\pm}+A_{H^\pm}+A_{\tilde\chi^\pm} \right|^2,
\label{eq_mueg}
\eeq
where the three contributions correspond to the diagrams a), b) and c) in Fig. \ref{1}, respectively. The LFV process in the first two contributions involves neutrinos in the loop. For the very small neutrino masses $\lesssim 1\,\mathrm{eV}$, as it follows from  eq.(\ref{SM-Br}),  one gets a negligibly small branching ratio. The same statement is valid for the charged Higgs boson whose mass is expected to be a few 100 GeV. Thus, one is actually left with the contribution of the sneutrino depicted in diagram c). 

The chargino contribution to $A_{\tilde\chi^\pm}$ is given by~\cite{Hisano:1996qq}
\begin{eqnarray}
\label{Acha}
A_{\tilde\chi^\pm} &=&\sum_{a=1}^2  \frac{m_W}{m_{\tilde\chi_a^\pm}}  \sum_{k=1}^3\left[\frac{m_W}{m_{\tilde\chi_a^\pm}}\left| U_{a1} \right|^2 U_{2k}^{\tilde\nu}U_{1k}^{\tilde\nu^*} \,f_{1}\left(\frac{m_{\tilde \nu_k}^2}{m_{\tilde \chi_a^\pm}^2}\right) \right.  
\nonumber\\
&&
\left. - \frac{1}{\sqrt 2\cos\beta}\, U_{a1} V_{a2}^* U_{2k}^{\tilde\nu} U_{1k}^{\tilde\nu^*} \,f_{2}\left(\frac{m_{\tilde \nu_k}^2}{m_{\tilde \chi_a^\pm}^2}\right)\right]
\eeqa
where the matrices $U$ and $V$ are the chargino mixing matrices and $U^{\tilde\nu}$ is the sneutrino mixing matrix. 
In our convention, the sneutrino flavour eigenstate basis is rotated to the sneutrino mass eigenstate basis in the same way, as it is in the neutrino sector: $\tilde\nu_{l_L} = \sum_{i} U^{\tilde\nu}_{li}\,\tilde\nu_{i_L}$.
In the limit of a minimal LFV the sneutrino mass matrix in eq. (\ref{Acha}) is represented by the PMNS matrix.
The functions $f_{1}(x)$ and $f_{2}(x)$ can be written as
\beqa
f_{1}(x)&=&\frac{1}{12\left(1-x\right)^4}\left( 2x + 3x^2 -6x^3 + x^4 + 6x^2\ln x \right),\\
f_{2}(x)&=&\frac{1}{2\left(1-x\right)^3}\left(1 - 4x + 3x^2 - 2x^2\ln x \right).
\eeqa
Note that in all the cases one has the product of the PMNS matrix elements $U_{\mu i}U_{e i}^*$ which being summed over $i$ gives zero due to the unitarity of the PMNS matrix. However, this product comes with the function $f\left(m_{\tilde \nu_k}^2/m_{\tilde \chi_j^\pm}^2\right)$ which depends on the mass of the i-th sneutrino. If all sneutrinos are degenerate, this function is universal and the result is zero. So the whole contribution crucially depends on the splitting in the sneutrino sector. In the MSSM with the universal boundary conditions the splitting is achieved via the non zero tau Yukawa coupling of the third generation. Hence, effectively, one has the contribution of the third generation
\beqa
A_{\tilde\chi^\pm} \sim U_{\mu 3}U_{e3}^* \left[f_{1,2}\left(\frac{m_{\tilde \nu_3}^2}{m_{\tilde \chi_j^\pm}^2}\right)-f_{1,2}\left(\frac{m_{\tilde\nu_1}^2}{m_{\tilde \chi_j^\pm}^2}\right)\right]\sim  \cos\theta_{13}\sin\theta_{13}\sin\theta_{23}.
\label{As13}
\eeqa
If the angle $\theta_{13}$ is zero, the whole contribution vanishes. If, on the contrary, it is big, the obtained decay width can contradict the experimental data, as it will be clear later. 
So the value of $\theta_{13}$ becomes crucial.
%

Since the muon decays to almost 100\% as $\mu\to e \bar \nu \nu$,  $\Gamma_{tot} =\Gamma(\mu \to e \bar\nu \nu)$, and
the branching ratio $Br(\mu \to e\gamma)$ can be written as
\beq
Br(\mu\to e \gamma)=\frac{\Gamma(\mu\to e \gamma)}{\Gamma(\mu \to e \bar\nu \nu)}
\label{Gmueg_over_Gmuenunu}
\eeq 
which finally gives, according to eqs. (\ref{wid}) and (\ref{eq_mueg}), 
\beq
Br(\mu \to e \gamma)=\frac{6\alpha}{\pi} \left|A_{\tilde\chi^\pm} \right|^2.
\eeq

%
At this point a question can arise whether and to which extent our results would be sensitive to the contribution of a hypothetical right-handed sneutrino which is not indeed taken into account in eq.(\ref{Acha}). 
In fact, a right-handed sneutrino does not belong to the particle content of the MSSM. Therefore, no mixing between left-handed and right-handed sneutrinos can occur and the sneutrino mass matrix is given by
\beqa
\mathcal{M}_{\tilde \nu}^2 =\left(
\begin{array}{cc}
m^2_{\tilde L}+\frac{1}{2}M_Z^2\cos 2\beta \, \mathds{1}_{3\times3} & 0_{3\times 3}  \\
0_{3\times 3} & 0_{3\times 3} \\
\end{array}
\right).
\eeqa
However, the MSSM can be extended by introducing right-handed neutrinos and their supersymetric partners. 
Further, with the implementation of a seesaw mechanism of type I \cite{Seesaw_type_1_1, Seesaw_type_1_2, Seesaw_type_1_3, Seesaw_type_1_4, Seesaw_type_1_5} the neutrino masses and mixing angles can be generated. 
In the presence of Majorana and Dirac mass terms in the Lagrangian the seesaw mechanism provides one light and one heavy neutrino mass eigenstate with masses $m_{\nu_\mathrm{light}} = m_D^2/M_R$ and $m_{\nu_\mathrm{heavy}}=M_R$, respectively, where higher order terms in $1/M_R$ are neglected, $m_D$ is the Dirac mass and $M_R\sim 10^{12}$ GeV is the Majorana mass. 
In this MSSM-seesaw model there is a contribution from the heavy neutrino mass eigenstate to the $W^-$ and $H^-$ loops and from the heavy sneutrino mass eigenstate to the $\tilde\chi^-$-loop. 
The contribution from the heavy neutrino to the LFV transitions is suppressed by a very small mixing angle $\theta^2 \approx m_D^2/M_R^2 = m_{\nu_{\mathrm{light}}}/M_R$ and, therefore, is completely negligible. 

Further, we can estimate the impact of the Majorana mass on the chargino mediated LFV process.
The Majorana mass enters into the LR, RL and RR blocks of the sneutrino mass matrix \cite{Grossman:1997is, Dedes:2007ef}.
Diagonalising the sneutrino mass matrix (see i.e. Ch. 2.1 in ref. \cite{Heinemeyer:2010eg}) one obtains corrections to the mass of the lighter sneutrino which are of the order of the light neutrino mass. The heavier sneutrino obtains a mass $m_{\tilde\nu_\mathrm{heavy}}=M_R$. Its contribution to the chargino mediated process is again suppressed by a very small mixing angle given above and, in addition, by the fact that the function $f_2$ vanishes for a big argument, $f_2(x\rightarrow \infty) \rightarrow 0$. Hence, the effect of the introduction of the Majorana masses is negligible and there is practically only the contribution from the left-handed sneutrinos to the chargino mediated $\mu \to e\gamma$ transition. The heavy sneutrinos decouple and the low-energy sneutrino mass eigenstates are dominated by the $\tilde \nu_L$ components.

Assuming mSUGRA universality conditions at the unification scale
the LL block of the sneutrino mass matrix can also be affected by the large mass $M_R$ through the renormalisation group running from the unification scale down to the SUSY scale. This leads to a LFV caused by renormalisation group effects.
The main idea is that the LL block of the sneutrino mass matrix is modified through a radiative correction to the soft slepton mass matrix $m^2_{\tilde L}$.
In leading logarithmic approximation the correction leads to off-diagonal elements in the sneutrino mass matrix which are typically smaller by a factor of $\sim10^{-4}$ in comparison with the diagonal ones.
Collecting all the flavour violation effects into the PMNS matrix, this additional flavour violation appears as a correction of the PMNS matrix elements of an order of $\sim 10^{-7}$. 
Hence, the radiative LFV does not lead to any sizeable effects in our case of study.
For a comprehensive study of the radiative LFV we refer to refs. \cite{Abada:2010kj, Deppisch:2010sv, Deppisch:2003wt, Deppisch:2002vz, Hisano:1995nq, Hisano:1995cp, Hisano:2009ae} and references therein. 

\section{Numerical analysis} 
We calculate the \mueg\ branching ratio in the MSSM with minimal LFV and mSUGRA universality conditions. 
The boundary conditions of this so-called constrained MSSM (CMSSM) imposed on the multidimensional MSSM parameter space imply that at the GUT scale all the sleptons, squarks and Higgs bosons have a common scalar mass $m_0$, all the gauginos are unified at the common gaugino mass $m_{1/2}$, and so all the tri-linear terms assume a common tri-linear mass parameter $A_0$. 
In addition, at the electroweak scale one selects the ratio of Higgs vacuum expectation values $ \tan \beta $ and sgn$(\mu)$, where $\mu$ is the higgsino mass parameter of the superpotential.
So one is left with the five-dimensional parameter space $(m_0, m_{1/2}, A_0, \tan \beta, sgn(\mu) )$.
For our numerical analysis we fix the value of $sgn(\mu)=1$ and the parameters $A_0$ and $\tan\beta$ vary for three different points $(m_0, m_{1/2})$. The dependence on $|A|$  comes from the RG equations for the running sneutrino masses.

We plot in Fig.(\ref{2}) on the left the relation between the branching ratio $Br(\mu \to e\gamma)$ and $A_0$ for different values of $\tan\beta$ for the points  $(m_0, m_{1/2})=(500, 500)\,\mathrm{GeV}$, $(m_0, m_{1/2})=(1500, 250)\,\mathrm{GeV}$ and  $(m_0, m_{1/2}) = (500, 900)\,\mathrm{GeV}$. 
The point $(m_0, m_{1/2})=(500, 500)\,\mathrm{GeV}$ has been chosen so that it is allowed by other processes, i.e., the branching ratios $B \rightarrow X_s \gamma$ and $ B \rightarrow l^+ l^- $, the anomalous magnetic moment of the muon as well as by experimental limits obtained by direct searches of the Higgs boson and Dark Matter in the Universe \cite{DK}. 
With the point $(m_0, m_{1/2})=(1500, 250)\,\mathrm{GeV}$ we can analyse a scenario with a heavy sneutrino and a light chargino; it corresponds to the beginning of the focus-point region. The choice $(m_0, m_{1/2})=(500, 900)\,\mathrm{GeV}$ represents the beginning of the so-called co-annihilation region  which is the opposite scenario. 
One should have in mind that values for $m_{1/2}\lesssim 250\, \mathrm{GeV}$ are excluded by direct Higgs searches and $B \rightarrow X_s \gamma$ even for small $\tan\beta$. 
On the other hand, the process $B\to\mu^+\mu^-$ is not compatible with small $m_0$.
We calculate the numerical values of the PMNS matrix elements with $\sin^2\theta_{13} = 0.016$ obtained by a global analysis of all available neutrino oscillation data \cite{Fogli:2008jx}.  

Our results show that for the points $(m_0, m_{1/2})=(500, 500)\,\mathrm{GeV}$ and  $(m_0, m_{1/2})=(500, 900)\,\mathrm{GeV}$ the branching ratio $Br(\mu\to e \gamma)$ grows with $A_0$ for all values of $\tan\beta$ while in the case of the point $(m_0, m_{1/2})=(1500, 250)\,\mathrm{GeV}$ we see the opposite trend. 
The current experimental upper bounds for $Br(\mu\to e \gamma)$ represented by two horizontal solid lines on the plots do not allow $\tan\beta\gtrsim 35$ even for a small $A_0$ and heavy chargino. An increasing $A_0$ together with a small chargino mass imposes even more stringent bound on the maximal value of the parameter $\tan\beta$.  

Note, however,  that these conclusions  are valid for the fixed value of $\sin^2\theta_{13} = 0.016$. At the same time, 
as was already demonstrated in eq. (\ref{As13}), our results crucially depend on the neutrino mixing parameter $\theta_{13}$. 
This proportionality is explicitly shown in the plots on the right side of Fig.(\ref{2}) where the value of $A_0$ is fixed, $A_0=0$, and we treat $\sin\theta_{13}$ as a free parameter.
The branching ratio $Br(\mu\to e \gamma)$ grows up with  $\theta_{13}$ and exceeds the experimental upper bounds even for small values of $\tan\beta$ if $\sin\theta_{13}$ is big enough. On the contrary, for small values of $\sin\theta_{13}$ all the values of $\tan\beta$ are allowed. 

\begin{figure}[htb]\vspace{0.3cm}
\begin{center}
 \epsfxsize=7.9cm
 \epsffile{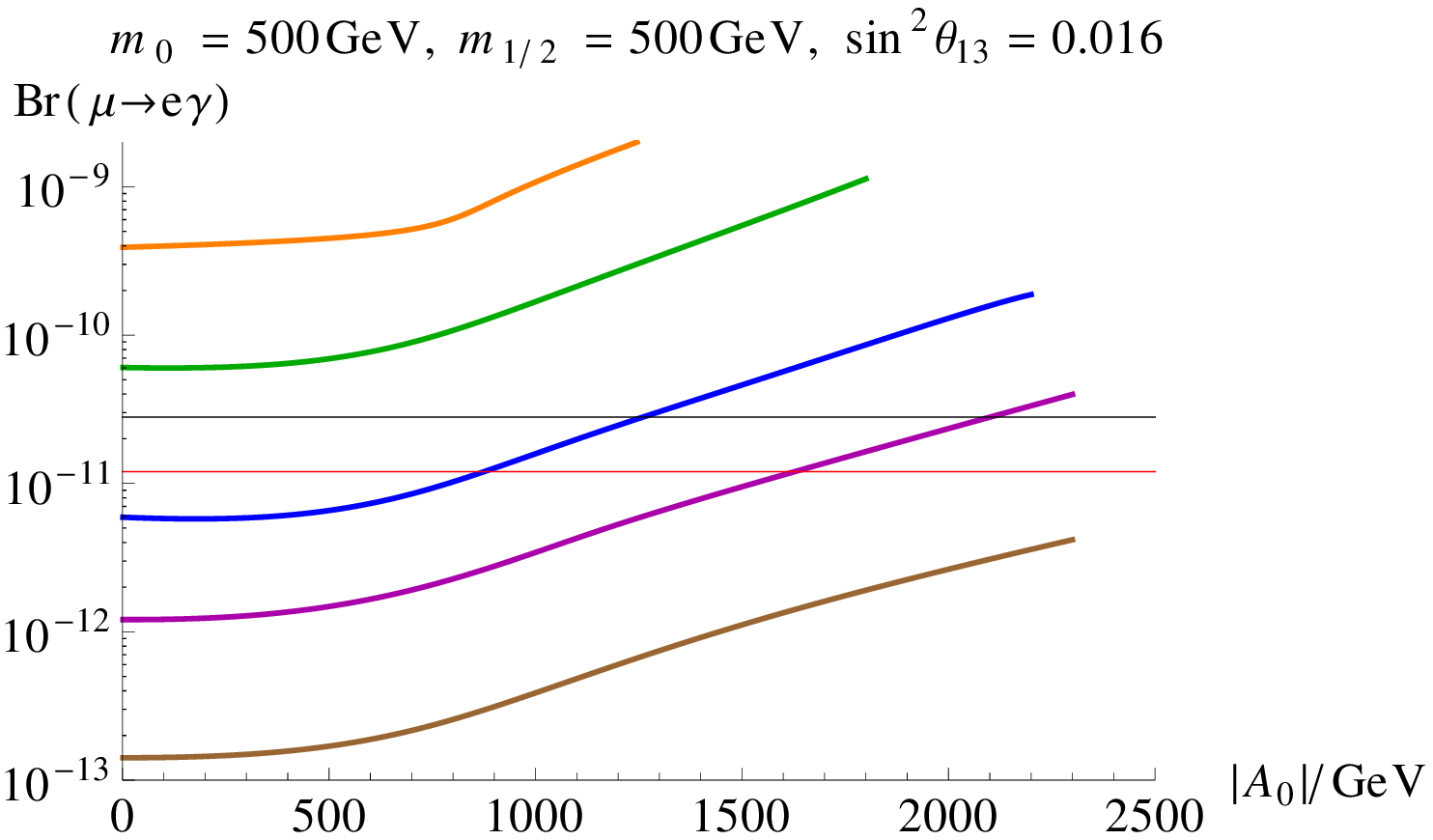}\hspace{0.0cm}\vspace{1cm}
 \epsfxsize=7.9cm
 \epsffile{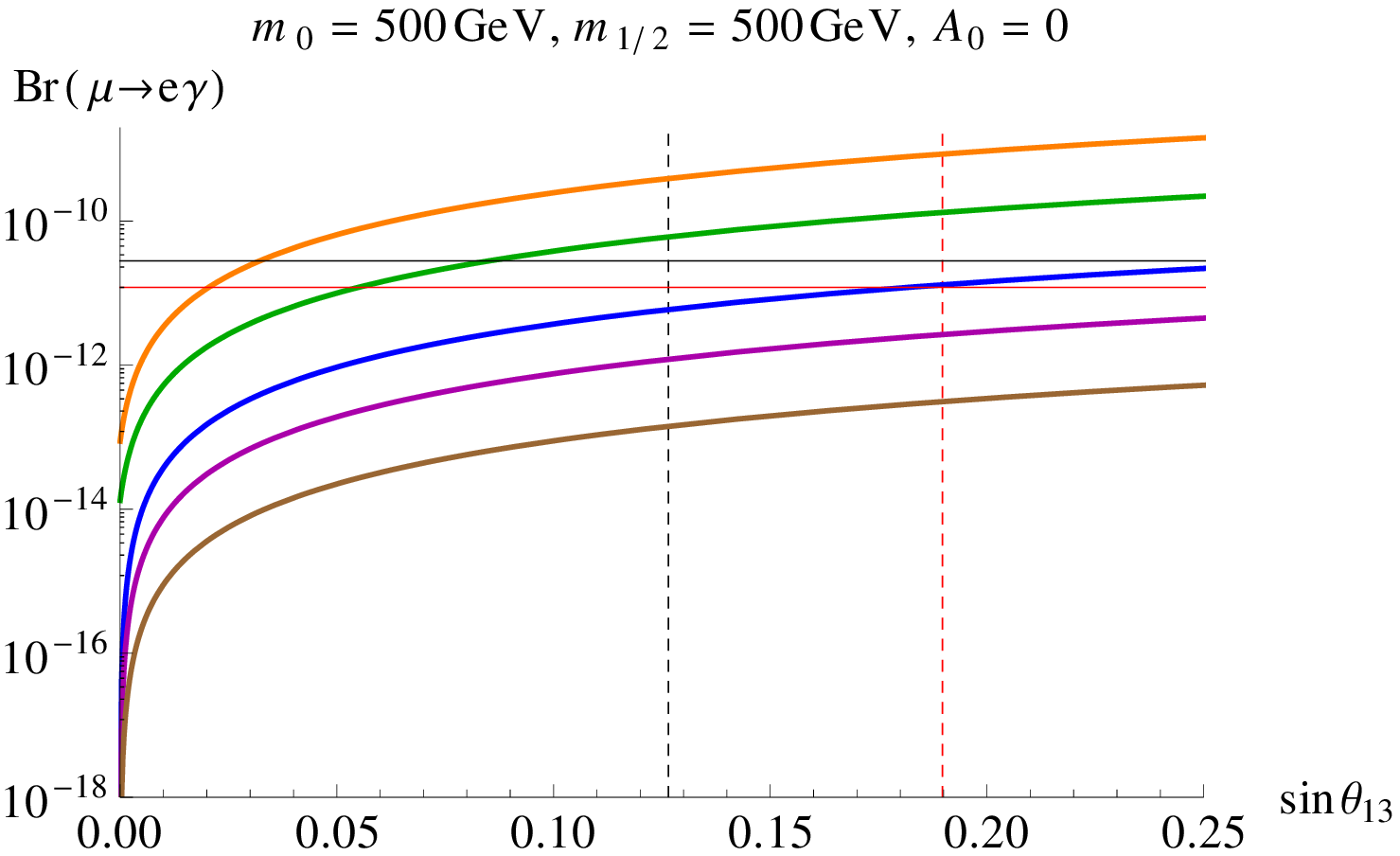}
 \epsfxsize=7.9cm
 \epsffile{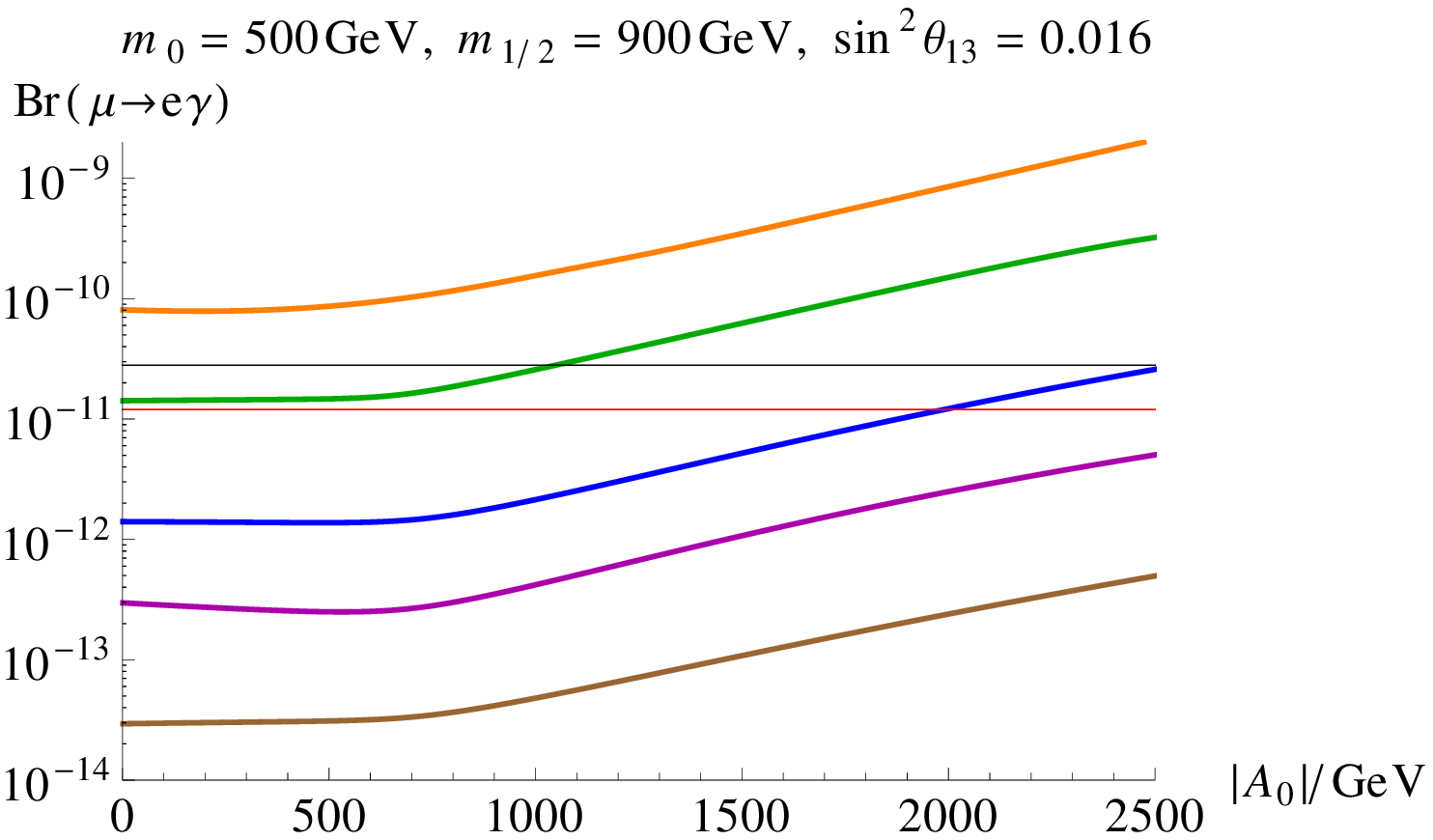}\hspace{0.0cm}\vspace{1cm}
 \epsfxsize=7.9cm
 \epsffile{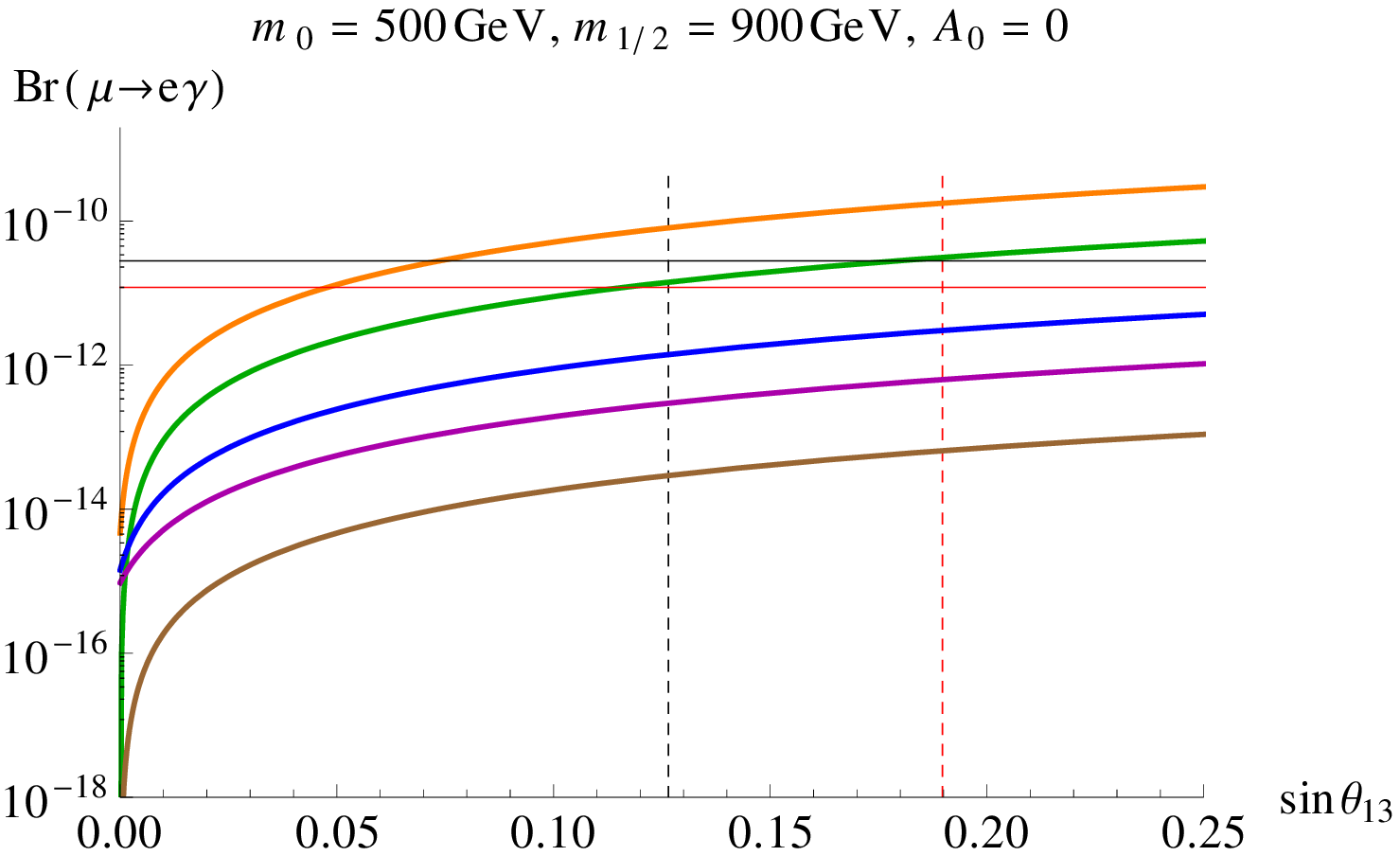}
 \epsfxsize=7.9cm
 \epsffile{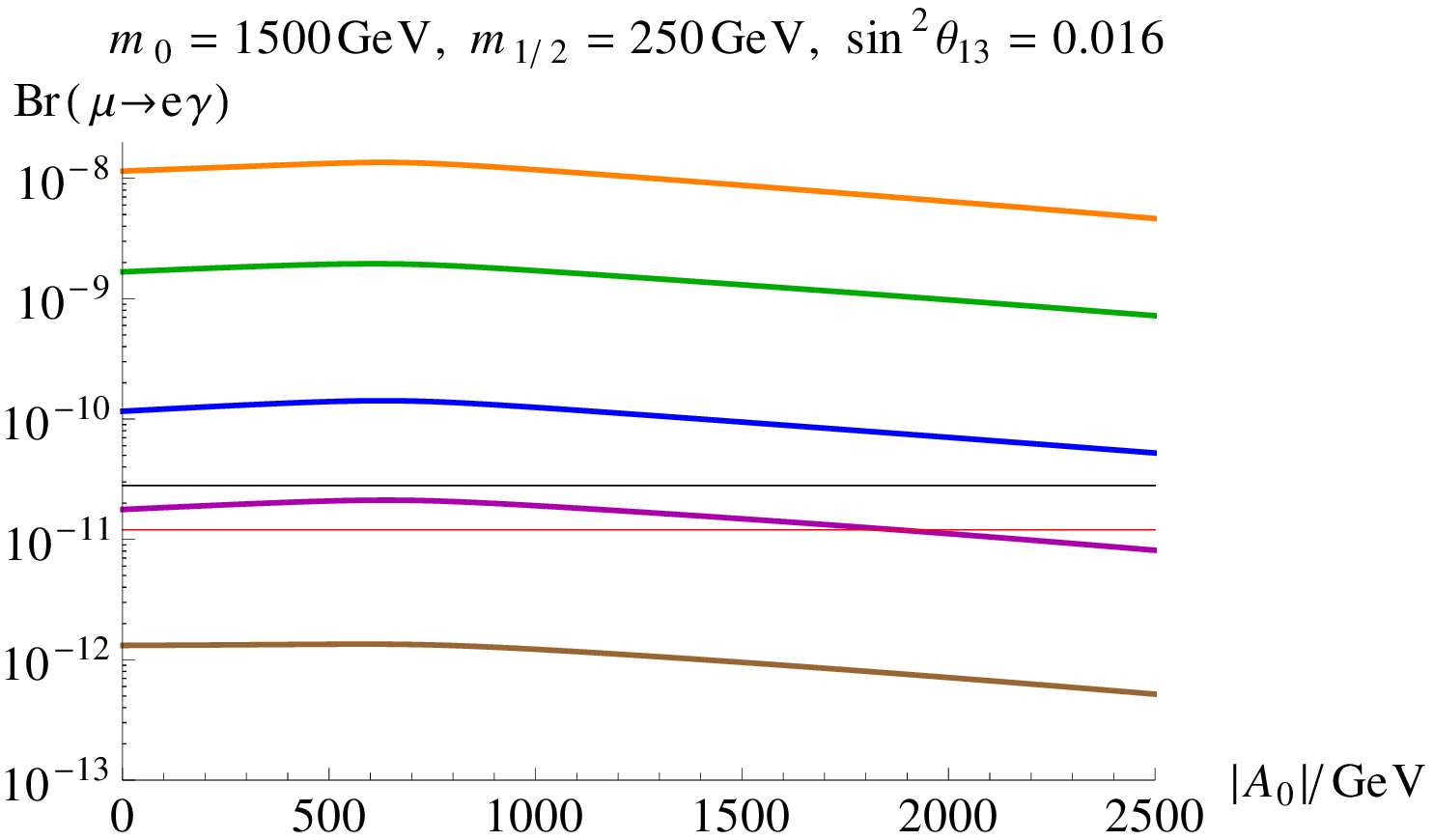}\hspace{0.0cm}\vspace{0.5cm}
 \epsfxsize=7.9cm
 \epsffile{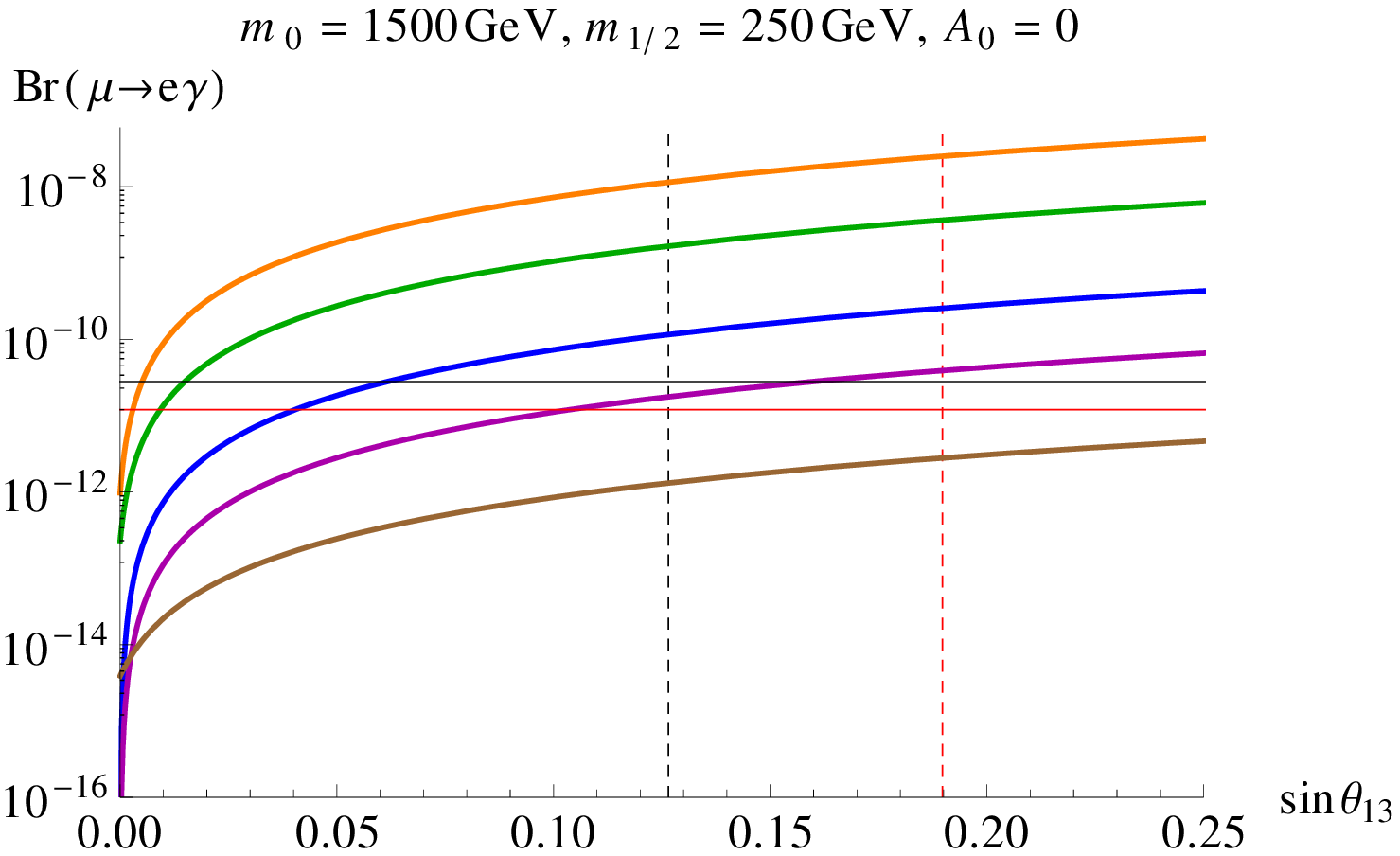}
 \caption{The branching ratio for the $\mu\to e\gamma$ decay 
as a function of $A_0$ and $\tan\beta$  (left) and as a function of $\sin\theta_{13}$ and $\tan\beta$ (right) for the points $(m_0, m_{1/2})=(500, 500)\,\mathrm{GeV}$, $(m_0, m_{1/2})=(1500, 250)\,\mathrm{GeV}$ and  $(m_0, m_{1/2}) = (500, 900)\,\mathrm{GeV}$, respectively. From top to bottom the curves correspond to $\tan\beta = 40, 30, 20, 15, 10$. The upper solid horizontal line corresponds to the experimental upper bound obtained by the MEG experiment $Br(\mu\to e \gamma)\leq 2.8\cdot10^{-11}$ \cite{Adam:2009ci} and the lower horizontal line to the one 
obtained by the MEGA experiment $Br(\mu\to e \gamma)\leq 1.2\cdot10^{-11}$ \cite{Brooks:1999pu}.  The left dashed vertical line on the plots on the right corresponds to the mean value of $\theta_{13}$ obtained by a global analysis of all available neutrino oscillation data $\sin^2\theta_{13} = 0.016$ \cite{Fogli:2008jx} and the right dashed vertical line represents the upper bound $\sin^2\theta_{13}\leq 0.056$ at 99.73\% C.L. \cite{Schwetz:2008er}. For all plots $sgn(\mu)=1$.}
 \label{2}
 \end{center}
 \end{figure}

\section{Discussion}

We have shown that the \mueg\ decay might serve as a direct manifestation of physics beyond the SM, in particular supersymmetry.
Experimental  bounds are very close to the predicted values. One has a unique combination of SUSY predictions with
the most involved measurement in neutrino physics, namely, the mixing between the first and the third generations $\theta_{13}$.
Provided the value of  $\theta_{13}$, the  \mueg\ decay would place  more stringent bounds on the MSSM parameter space than other rare decays.   One can see that the high $\tan\beta$ scenario of the MSSM favoured by dark matter abundance~\cite{deboer}
might contradict the \mueg\ decay for the essential part of the bulk region if the value of $\theta_{13}$ is big enough. On the contrary,
global analysis of the SUSY parameter space surely prefers small values of  $\theta_{13}$. It seems that the resolution of both puzzles might come together.

\section*{Acknowledgements}
We are grateful to S.M.Bilenky, M.V.Danilov, J.A.Budagov and M.I.Vysotsky for stimulating discussions.
Financial support
from RFBR grant \# 11-02-01177 and the Ministry of Education and
Science of the Russian Federation grant \# 1027.2008.2 is kindly
acknowledged.

\end{document}